\documentclass[12pt]{article}
\usepackage{setspace}
\usepackage{amsmath}
\usepackage{graphicx}
\usepackage{cases}
\usepackage{subfigure}
\usepackage{natbib}
\usepackage{psfrag}
\usepackage{amssymb}
\usepackage{color,graphics,epsfig}
\usepackage{appendix}
\usepackage{fancyhdr}
\usepackage{xr}
\def\be{\begin{equation}}
\def\ee{\end{equation}}
\def\bea{\begin{eqnarray}}
\def\eea{\end{eqnarray}}

\def\noi{\noindent}

\begin{document}

\title{Interference effects of deleterious and beneficial mutations in large asexual populations}

\author{Kavita Jain \\\mbox{}\\
Theoretical Sciences Unit, \\Jawaharlal Nehru Centre for Advanced Scientific Research, \\Jakkur P.O., Bangalore 560064, India }

\maketitle

\newpage

\noindent
Running head: Fixation of beneficial mutations
\bigskip

\bigskip

\noindent
Correspondence:
\texttt{jain@jncasr.ac.in}

\bigskip
   
\noindent
\textbf{Abstract:} Linked beneficial and deleterious mutations are known to decrease the fixation probability of a favorable mutation in large asexual populations. While the hindering effect of strongly deleterious mutations on adaptive evolution  
has been well studied, how weak deleterious mutations, either in isolation or with superior beneficial mutations, influence the fixation of a beneficial mutation has not been fully explored. 
Here, using a multitype branching process, we obtain an accurate analytical expression for the fixation probability when deleterious effects are weak, and exploit this result along with the clonal interference theory to investigate the joint effect of linked beneficial and deleterious mutations on the rate of adaptation. We find that when the mutation rate is increased beyond the beneficial fitness effect, the fixation probability of the beneficial mutant decreases from Haldane's classical result towards zero. This has the consequence that above a critical mutation rate that may depend on the population size, 
the adaptation rate decreases exponentially with the mutation rate and is independent of the population size. In addition, we find that for a range of mutation rates, both beneficial and deleterious mutations interfere and impede the adaptation process in large populations. We also study the evolution of mutation rates in adapting asexual populations, and conclude  that linked beneficial mutations have a stronger influence on mutator fixation than the deleterious mutations. 
 
\newpage


Adaptation is driven by beneficial mutations. In a large and fully recombining population, the chance that a beneficial mutation fixes is two times its selective effect \citep{Haldane:1927}.  But in an asexual population, this fixation probability is reduced either when the beneficial mutant  arises on a genetic background with many deleterious mutations ({\it background selection}; 
 \citet{Charlesworth:1994}), because competing favorable mutations are present ({\it clonal interference}; \citet{Gerrish:1998}), or, in general, due to both reasons.
 
The rate of adaptation in large asexual populations has been a subject of a number of investigations, and recent experimental \citep{Lang:2013,Wiser:2013,Batista:2014,Levy:2015} and theoretical studies \citep{Gerrish:1998,Orr:2000b,Wilke:2004,Desai:2007a,Park:2007} have focused on understanding the effect of competition between linked beneficial mutations  on adaptation dynamics \citep{Sniegowski:2010},  
but these work neglect the influence of linked {\it deleterious} mutations. This is justified when 
mutation rates are small or selection against deleterious mutations is strong 
so that the beneficial mutation arises on the deleterious mutation-free genetic background \citep{Charlesworth:2012}. 

However, mutation rates can rise by several orders of magnitude, at least, in adapting populations in laboratory \citep{Raynes:2014}, and deleterious effects can be small or moderately sized \citep{Eyrewalker:2007}. In these scenarios, linked deleterious mutations can decrease the  fixation probability of beneficial mutations \citep{Johnson:2002},   and therefore the rate of adaptation. In addition, superior beneficial mutations may also 
interfere with the spread of a beneficial mutation in sufficiently large populations, and have a further negative impact on adaptation rate \citep{Jiang:2011,Penisson:2017}. However, to the best of our knowledge, closed expressions for the fixation probability and the extent to which linked mutations undermine adaptive evolution have not been obtained when deleterious effects are weak.  
 
The above considerations may also have a bearing on the evolution of mutation rates.  
In a recent work, \citet{Raynes:2018} have experimentally shown that asexual mutators are favored in populations larger than a critical population size, and argue that this critical value is inversely proportional to the fixation probability of a beneficial mutation. Their theoretical analysis, however, neglects any linkage effects.  But as linked mutations decrease the fixation probability, it is important to ask how large the population size should be for the mutators to survive in general settings. 

To address these questions, we consider a general model in which the beneficial mutant may either have the same mutation rate as that of the population in which it arises  or it may be a mutator, 
and find an analytical expression for the fixation probability when deleterious effects are weak. 
We then use this result along with the clonal interference theory \citep{Gerrish:1998} to understand the joint  effect of linked deleterious and beneficial mutations on the rate of adaptation. We identify regions in the space of population size and mutation rates where either beneficial or deleterious or both kind of mutations interfere; these results are summarized in Fig.~\ref{figphase}. We also investigate the relationship between population size and mutation rates in adapting populations, and find that clonal interference strongly affects the critical population size above which mutators are favored (see, Fig.~\ref{fig7}). \\

\noi{\bf \large{Models and Methods}}

 \noi We consider a haploid asexual population of finite size $N$. In every generation, an individual acquires deleterious mutations that are Poisson-distributed with mean $u_d$, and each such mutation decreases its fitness by a factor $1-s_d$.  We work with populations of size $N \gg s_d^{-1} e^{u_d/s_d}$ in which the Muller's ratchet  \citep{Muller:1964,Jain:2008b} operates slowly and the population stays close to the deterministic mutation-selection equilibrium before the first click of the ratchet. 
In this resident population, beneficial mutations that are also Poisson-distributed arise at rate $U_b$, and each beneficial mutation increases the fitness  of the individual by a factor $1+s_b$. Beneficial mutants also acquire further deleterious mutations  at rate $U_d$ that decreases their fitness by $1-s_d$ per deleterious mutation. The deleterious effects are assumed to have a fixed size $s_d$, but the beneficial fitnesses are chosen from a truncated exponential distribution with mean ${\bar s_b}$ \citep{Eyrewalker:2007}. 

In this article, we are mainly interested in understanding how {\it weak deleterious mutations} (as defined in the next section) affect the fixation of beneficial mutations. We will consider the situation when the beneficial mutant is under direct selection ($U_d=u_d, U_b=u_b$), and analyze how linked beneficial and deleterious mutations impact the rate of adaptation in asexuals.  Motivated by a recent work \citep{Raynes:2018}, we also study the case in which the invading mutant is a mutator with mutation rates $U_d > u_d$ and $U_b=u_b (U_d/u_d)$, and address how interference by linked mutations affect the evolution of mutation rates in adapting asexual populations. 

Using a multitype branching process \citep{Harris:1963,Patwa:2008} and some ideas developed in \citet{Jain:2017b}, we first obtain analytical expressions for the fixation probability of a beneficial mutant of fixed effect size in the general model described above. We then use this result and the clonal interference theory \citep{Gerrish:1998} to study the combined influence of deleterious and beneficial mutations. 
Since we work in the parameter region where the deleterious effects are small, large population sizes are required to ensure that Muller's ratchet clicks slowly. This  limits the range of parameters that can be investigated numerically, but some results from individual-based simulations of the above model are reported in Supplemental Material, File~{S1} for $N \sim 10^5$, and found to agree reasonably well with the theory developed in the main text. \\

\noi{\bf \large{Fixation Probability of a Beneficial Mutant}}

\noi We now proceed to calculate the fixation probability of a single beneficial mutant in a large resident population using the standard multitype branching process. As detailed in Appendix~\ref{app_quad}, the fixation probability $p_i$ of a mutant arising in 
a stationary genetic background with $i$ deleterious mutations obeys the following recursion equation, 
\bea
1-p_i = \exp \left[-(1+s_b) (1-s_d)^i e^{u_d} \sum_{j=0}^{\lambda-i}  \frac{e^{-U_d} U_d^j}{j!} p_{i+j}  \right]~,~ 0 \leq i \leq \lambda ~.~
\label{extprob1} 
\eea
In the above equation, the nonnegative integer $\lambda$ is the maximum number of deleterious mutations that the beneficial mutant can carry in order to have a nonzero fixation probability, and is given by 
\be
\lambda =\left  \lfloor \frac{\ln(1+s_b)-(U_d-u_d)}{-\ln(1-s_d)} \right \rfloor ~,~1+s_b > e^{U_d- u_d}~,~
\label{imaxf}
\ee
where $\lfloor x \rfloor$ denotes the maximum integer less than or equal to $x$. 
For $U_d=u_d$, our (\ref{extprob1}) and (\ref{imaxf}), respectively, reduce to (12) and (10) of \citet{Johnson:2002}. The total fixation probability is obtained by averaging over all the genetic backgrounds, 
\be
P = \sum_{i=0}^{\lambda} p_i f_i~,~
\label{fptot}
\ee  
where the probability $f_i$ that the beneficial mutation arises in a genetic sequence with $i$ deleterious mutations is 
given by \citep{Kimura:1966,Haigh:1978} 
\be
f_i = e^{-\mu} \frac{\mu^i}{i !} 
\label{poi}
\ee
and the average number of deleterious mutations $\mu=u_d/s_d$. Using the boundary condition $p_{\lambda+1}=0$, one can solve (\ref{extprob1}) for the fixation probability numerically \citep{Johnson:2002}. 

However, it is possible to make analytical progress if one exploits the fact that besides the fixation probability $p_i$, all the mutation rates and selection coefficients are also smaller than one \citep{Jain:2017b}. On expanding (\ref{extprob1}) in a power series about these small quantities and retaining terms to quadratic orders (see Appendix~\ref{app_quad} for details), we obtain the following equation for the fixation probability, 
\be
\frac{p_i^2}{2} \approx U_d p_{i+1}+ (s_b-i s_d+u_d-U_d) p_i~,~0 \leq i \leq \lambda~,~
\label{receqn}
\ee
where 
\be
\lambda \approx \lfloor \frac{s_b-U_d+u_d}{s_d} \rfloor ~,~s_b > U_d- u_d~.~
\label{imax}
\ee
Note that the approximations employed here to arrive at (\ref{receqn}) are similar to those used to obtain Haldane's 
classical result \citep{Haldane:1927} for fixation in a single genetic background.  

Figure~\ref{fig1} shows the fixation probability $p_i$ in the genetic background with $i$ deleterious mutations, and we find that the quadratic approximation (\ref{receqn}) somewhat overestimates the exact fixation probability (\ref{extprob1}).  However, as (\ref{receqn}) enables one to make analytical progress, we employ the quadratic approximation for the rest of the article. 

\noi{\bf When deleterious effects are strong}

\noi When the effect of deleterious mutations is large enough, the beneficial mutant can fix in only a few genetic backgrounds. In particular, only the deleterious mutation-free background matters ($\lambda=0$) when $s_d > s_b- U_d+u_d > 0$. Then, using (\ref{receqn}), we immediately obtain $p_i=2 (s_b-U_d+u_d) \delta_{i,0}$, and
\be
P= \sum_{i=0}^{\lambda} f_i p_i=2 s_b \left(1-\frac{U_d-u_d}{s_b} \right) e^{-u_d/s_d}
\label{M2}
\ee 
on using (\ref{poi}). When $U_d=u_d$, the above equation reduces to the well known result for the fixation probability of a mutant under direct selection \citep{Charlesworth:1994,Peck:1994}. 

Equation (\ref{M2}) shows that as the mutator's strength $\sigma=U_d/u_d$ (for a given $u_d$) increases, its chance of fixation decreases since the mutant accumulates more deleterious mutations. The fixation of strong mutators has been studied ignoring this fact in \citet{Raynes:2018}, and we discuss this point further in a later section. With increasing $u_d$, both the resident and the mutant carry larger number of deleterious mutations and therefore the fixation probability again decreases. 
However, increasing the fitness advantage $s_b$ of the mutant 
enhances its chance of fixation; similarly, increasing the fitness cost $s_d$ of deleterious mutation means that the resident population and the mutant carry fewer deleterious mutations, and increases the fixation probability. 
Thus, as intuitively expected, the fixation probability increases with the selection coefficients $s_b$ and $s_d$, but decreases with increasing mutation rates $U_d$ and $u_d$. As we shall see below, not all of these qualitative trends continue to hold when the deleterious effects are weak. 

\noi{\bf When deleterious effects are weak}

\noi For a given $\lambda$, the definition (\ref{imax}) means that $\lambda \leq (s_b-U_d+u_d)/s_d < \lambda+1$. But for weak deleterious effects that correspond to large $\lambda$, we may write $\lambda \approx (s_b-U_d+u_d)/s_d$, and the recursion equation (\ref{receqn}) as
\be
\frac{p_i}{2} - s_d (\lambda-i) = U_d \frac{p_{i+1}}{p_i}~.~
\label{receqn2}
\ee

For weak deleterious effects, although the beneficial mutant can survive in $\lambda \gg 1$ genetic backgrounds, as the background frequency $f_i$ is Poisson-distributed with mean and variance equal to $\mu$ (see (\ref{poi})), the beneficial mutant is likely to arrive in those individuals that carry deleterious mutations lying between $\mu-\sqrt{2 \mu}$ and $\mu+\sqrt{2 \mu}$. Then, if $ \lambda \ll \mu-\sqrt{2 \mu}$, the beneficial mutant arrives in those genetic backgrounds in which it can not survive and hence its fixation probability is zero. On the other hand, if $ \lambda \gg \mu+\sqrt{2 \mu}$, the arrival of beneficial mutant occurs in those genetic backgrounds where it has a substantial chance of spreading, and the fixation probability is therefore expected to be nonzero. This simple {\it arrival-survival argument} suggests a nontrivial transition in the behavior of total fixation probability $P$: it is nonzero when the mutation rate $U_d \ll s_b - \sqrt{2 u_d s_d}$ and zero for $U_d \gg s_b + \sqrt{2 u_d s_d}$. This expectation is indeed borne out by our detailed analysis, as discussed below.  

\noi{\bf Bounds on fixation probability:} Since the probability $p_i$ is nonnegative and decreases monotonically with increasing number of background mutations $i$ (see Fig.~\ref{fig1}), the right-hand side (RHS) of (\ref{receqn2}) lies between zero and $U_d$. This implies that 
\be
2 s_d (\lambda-i ) < p_i <  2 (s_b+u_d-i s_d)~.
\label{bdpi}
\ee
These bounds have a simple meaning: the lower bound is obtained on assuming that the beneficial mutant can mutate but does not fix on any background other than the one it arose on, while on setting the deleterious mutation rate of the mutant to zero, the upper bound given by twice the relative fitness of the mutant with respect to the average fitness of the resident population \citep{Haldane:1927} is obtained. Figure~\ref{fig1} shows that the upper bound is a good approximation to the fixation probability obeying (\ref{receqn}) in genetic backgrounds with few deleterious mutations, whereas the lower bound works fairly well in backgrounds with many deleterious mutations. 

Although exact expressions for the bounds on total fixation probability  can be found by carrying out the sum in (\ref{fptot}), they are not particularly illuminating.  However, for large $\mu$, the Poisson distribution $f_i$ can be approximated by a Gaussian with mean and variance $\mu$, and the sum in (\ref{fptot}) by an integral. On performing the integrals, we find that for $\lambda, \mu \gg1 $, the bounds on the fixation probability are given by 
\be 
\sqrt{\frac{2 u_d s_d}{\pi}}  \left[ r \sqrt{\pi} (1+\textrm{erf}(r))+e^{-r^2} \right] < P < s_b \left[1+\textrm{erf}(r) \right]+ \sqrt{\frac{2 u_d s_d}{\pi}} e^{-r^2}~,~
\label{pbd}
\ee
where $r=(s_b-U_d)/\sqrt{2 u_d s_d}$ and $\textrm{erf}(r)=(2/\sqrt{\pi}) \int_0^r dx~e^{-x^2}$ is the error function which has  the property that $\textrm{erf}(r \to \pm \infty) \to \pm 1$. 
Then it is easy to see that in the limit $s_d \to 0$, there is a transition in the behavior of the above bounds at $U_d=s_b$.  
Specifically, the lower bound on the total fixation probability decreases linearly with $U_d$ as $2 (s_b - U_d)$ for $U_d < s_b$ and is zero for $U_d \geq s_b$.  The upper bound, on the other hand, equals $2 s_b$ for all  $U_d \leq s_b$ but vanishes for  $U_d > s_b$. 


\noi{\bf Fixation probability for $U_d/s_b \to 0$:} 
As mentioned above, the upper bound  on fixation probability is obtained by ignoring the mutations accumulated by the mutant. Taking these mutations into account for $u_d, U_d \to 0$ and $s_d \ll s_b$, we find that 
the relative fixation probability decreases as (see Appendix~\ref{app_pert})
\be
\frac{P}{2 s_b} \approx 1- \frac{U_d}{s_b} \frac{s_d}{s_b}~,~U_d \ll s_b~.~
\label{pert}
\ee


\noi{\bf Fixation probability for finite $U_d/s_b$:} To investigate the transition in the behavior of the fixation probability, we now focus on the parameter regime where $U_d/s_b$ is finite. In Appendix~\ref{app_finp}, we show that for large $\lambda$ and $\mu$, the fixation probability 
\be
p_i \approx 2 s_d (\lambda-i)+ 2 U_d \left(\frac{s_d}{U_d}\right)^{2^{i+1-\lambda}} ~.
\label{pexpr}
\ee
The above expression interpolates between the bounds in (\ref{bdpi}): The first term on the RHS, which decreases linearly with increasing number of deleterious mutations,  
is simply the lower bound on the probability $p_i$. The second term which is of a double-exponential form ({\it i.e.,} $e^{-e^{-x}}$) approaches a constant $2 U_d$ for $i \ll \lambda$, but decreases quickly to $2 s_d$ for $i \sim \lambda$. As a result, (\ref{pexpr}) matches with the upper bound in (\ref{bdpi}) when the background has few deleterious mutations (see, Fig.~\ref{fig1}). The result (\ref{pexpr}) is also compared with the numerical solution of (\ref{receqn}) in Fig.~\ref{fig1}, and we find a good agreement.  

As described in Appendix~\ref{app_finp}, for large $\lambda$ and $\mu$, (\ref{pexpr}) yields the total fixation probability to be 
\be
P \approx s_b \left[1+\textrm{erf}(r) \right]+ \sqrt{\frac{2 u_d s_d}{\pi}} e^{-r^2} + U_d \left[ \textrm{erf} ({\tilde r}) -\textrm{erf}(r)\right]~,~
\label{fullpexpr}
\ee
where 
\bea
r  &=& \frac{s_b-U_d}{\sqrt{2 u_d s_d}} ~,~\\
{\tilde r} &=& r - \frac{1}{\sqrt{2 \mu}} \left[\frac{1}{\ln 2} \ln \left(\frac{\ln (U_d/s_d)}{\ln \sqrt{2}} \right)\right] ~.
\eea
Figure~\ref{fig2} shows a comparison between the total fixation probability obtained by using the numerical solution of  quadratic approximation (\ref{receqn}) and the analytical expression (\ref{fullpexpr}), and indicates that (\ref{fullpexpr}) overestimates the former for small $U_d/s_b$ and underestimates it for larger values. However, the difference is quite small, and as discussed in Appendix~\ref{app_finp} and shown in the inset of Fig.~\ref{fig2}, it decreases as the deleterious effect gets weaker. 

Figure~\ref{fig2} also shows that the upper bound (\ref{pbd}) grossly overestimates the total fixation probability (\ref{fullpexpr}) for high mutation rate. As discussed earlier, the upper bound captures the effect of the deleterious mutations that the beneficial mutant inherits from the genetic background on which it arose \citep{Charlesworth:1994}, but ignores the  
mutations accumulated by the beneficial mutant during evolution \citep{Penisson:2017}. The latter effect is taken care of by the last term on the RHS of (\ref{fullpexpr}) which is negative since the error function is an increasing function. 

Equation (\ref{fullpexpr}) shows that in the limit $s_d \to 0$, as $\textrm{erf}(r \to \pm \infty) \to \pm 1$, the total fixation probability equals $2 s_b$ for all $U_d \leq s_b$ and zero for $U_d > s_b$ which means that for very weak selection against deleterious mutations, a beneficial mutant does not spread if its mutation rate exceeds its beneficial effect but has the same chance of fixation as in a fully recombining population for {\it any} mutation rate below its beneficial fitness effect. For finite but small $s_d$, the  total fixation probability does not change sharply but decreases gradually from $2 s_b$ to zero when mutation rate $U_d$ lies between $s_b-\sqrt{2 u_d s_d}$ and $s_b+\sqrt{2 u_d s_d}$, as argued at the beginning of this section. 
Thus, as supported by Fig.~\ref{fig2}, for weak deleterious effects, the total fixation probability is well approximated by $2 s_b$ \citep{Haldane:1927} so long as the mutation rate is sufficiently small. But with increasing mutation rate, the  fixation probability decreases and eventually vanishes. \\

\noi{\bf \large{Interference Effects in Large Asexual Populations}}

\noi In this section, we set $U_d=u_d$ to make contact with previous work, and discuss how weak deleterious effects impact the rate of adaptation. 


\noi{\bf Fixation probability} 

\noi Equation (\ref{fullpexpr}) shows that for $U_d=u_d$, the  probability $P/(2 s_b)$ is a function of the scaled parameters $\lambda=s_b/s_d$ and $\mu=u_d/s_d$ \citep{Johnson:2002}, and decreases from one to zero with increasing $u_d$. Although one expects the fixation probability to decrease with decreasing $s_d$ also, the data in Fig.~\ref{fig3} paints a more complex picture: when $u_d > s_b$, the fixation probability decreases continuously from $2 s_b e^{-u_d/s_d}$ (see (\ref{M2})) to zero with decreasing deleterious effect; this is because the beneficial mutant is likely to accumulate more deleterious mutations as the selection against deleterious mutation gets weaker. 
But for $u_d < s_b$, the fixation probability $P$ is a U-shaped function: it is close to $2 s_b$ for $s_d \gg s_b$ \citep{Haldane:1927}, but as a consequence of the transition discussed in the last section, it is given by $2 s_b$ for $s_d \ll s_b$ also.   
For the parameters in Fig.~\ref{fig3}, the minimum in total fixation probability occurs at ${\lambda}^{-1}=s_d/s_b \approx 0.3$. But as our result (\ref{fullpexpr}) is valid for large $\lambda, \mu$ (refer Appendix~\ref{app_finp}), we are unable to provide an analytical estimate for the deleterious effect that minimizes the fixation probability. For the same reason, the results in Fig.~\ref{fig3} for the total fixation probability obtained by numerically iterating (\ref{receqn}) and the analytical expression (\ref{fullpexpr}) do not agree well when $\lambda$ and $\mu$ are small. 

The above discussion thus indicates that when the selection against deleterious mutations is weak, the loss of beneficial mutation can be mitigated if the mutation rate is small enough. To put in another way, large enough beneficial effect can overcome the deleterious effect of mutations; indeed, as the  inset of Fig.~\ref{fig3} shows, with increasing beneficial effect, the total fixation probability rises from zero towards $2 s_b$ when $s_b$ exceeds $u_d$. 
As we shall see below, this transition in the fixation probability plays an important role in our understanding of adaptation dynamics. 


\noi{\bf Substitution rate} 

\noi In a large population of size $N$, if the  beneficial mutations are produced at rate $u_b$ per individual and the beneficial fitnesses are distributed according to the probability distribution $\rho(s_b)$, the  expected substitution rate of beneficial mutations  is given by 
\be
E[k_b]=N u_b \int_0^{S_b} ds_b \rho(s_b) P(s_b) e^{-I(s_b)} ~.
\label{subsdef}
\ee
In the above equation, the effect of interfering beneficial mutations is captured by the factor $e^{-I(s_b)}$, where $I(s_b)=N u_b \ln (N s_b) s_b^{-1} \int_{s_b}^{S_b} ds'  \rho(s') P(s') $ is the average number of contending mutations with beneficial effect greater than $s_b$ 
that escape stochastic loss \citep{Gerrish:1998,Penisson:2017}, and the distribution $\rho(s_b)$ is a truncated exponential distribution with  mean ${\bar s}_b$ and upper bound $S_b \ll 1$. The latter condition is required because our results for fixation probability are derived by assuming that the selection coefficients are smaller than unity; however, it is a good approximation to replace the upper limit by infinity if 
${\bar s}_b \ll S_b$ (see below). 

The inset of Fig.~\ref{fig3} suggests that for $s_d \to 0$, we may approximate the total fixation probability by a step function at $s_b=u_d$ so that $P=2 s_b$ for $s_b > u_d$ and zero otherwise. This results in 
\be
\frac{E[k_b]}{E_0[k_b]} = \int_{u_d/{\bar s_b}}^{\infty} dx~x e^{-x} e^{-2 N u_b \ln(N {\bar s}_b x) (1+x^{-1}) e^{-x}} ~,
\label{Eks00}
\ee
where, $E_0[k_b]=N u_b \int_0^\infty ds_b \rho(s_b) 2 s_b =2 N u_b {\bar s}_b$ is the expected substitution rate in the absence of any linked mutations. The integral in (\ref{Eks00}) can be numerically evaluated (see Fig.~\ref{fig4}), but one can obtain an insight into its behavior by approximating it by $\int_{u_d/{\bar s_b}}^{\infty} dx~x e^{-x} e^{-{\tilde A} e^{-x}}$ \citep{Wilke:2004}, and arrive at
\begin{subnumcases}{\frac{E[k_b]}{E_0[k_b]} \approx}
\left(1+\frac{u_d}{{\bar s}_b} \right) e^{-{u_d}/{{\bar s}_b}} ,&${\tilde A} e^{-{u_d}/{{\bar s}_b}} \ll 1$ \label{subsN}\\
\frac{1}{{\tilde A}}\left({\ln {\tilde A}}-\frac{u_d}{{\bar s}_b} e^{-{\tilde A} e^{-{u_d}/{{\bar s}_b}}} \right),&${\tilde A} e^{-{u_d}/{{\bar s}_b}} \gg1, {{\bar s}_b}\ll {u_d}$ \label{submN} \\
\frac{{\ln {\tilde A}}}{{\tilde A}},&${\tilde A} e^{-{u_d}/{{\bar s}_b}} \gg1, {\bar s}_b \gg u_d$ \label{sublN} 
\end{subnumcases}
where ${\tilde A}=2 N u_b \ln (N {\bar s}_b)$. The condition ${\tilde A} e^{-{u_d}/{{\bar s}_b}}=1$ separates the regions where interference by linked beneficial mutations can be ignored and where it matters, and $u_d/{\bar s}_b=1$ gives the corresponding condition for deleterious mutations.  

Equation (\ref{subsN}) captures the effect of interference by deleterious mutations alone and shows that  for  ${\bar s}_b \ll u_d$, the substitution rate $E[k_b]$ decreases exponentially with increasing mutation rate, but for ${\bar s}_b \gg u_d$, there is no interference by either mutations as $E[k_b] \approx E_0[k_b]$. This behavior is, of course, a consequence of the transition in the fixation probability which vanishes for ${s}_b < u_d$ and equals $2 s_b$ for ${s}_b > u_d$. Equation (\ref{sublN}) is the substitution rate when only beneficial mutations interfere, and matches the known results in the clonal interference regime when deleterious effects are strong \citep{Wilke:2004}. The result (\ref{submN}) holds when both kind of linked mutations interfere, and is obviously smaller than (\ref{subsN}) and (\ref{sublN}). 
 
Figure~\ref{fig4} shows the scaled substitution rate as a function of the mean beneficial effect ${\bar s}_b$ for two population sizes, and we find that the result (\ref{Eks00}) for $s_d \to 0$ is a good approximation for the substitution rate when the deleterious effect is small but finite, and therefore we may analyze the behavior of $E[k_b]$ using (\ref{Eks00}). For the parameters in Fig.~\ref{fig4} and the population size $N=10^8$, clonal interference can be ignored since ${\tilde A}e^{-{u_d}/{{\bar s}_b}} <1$ for any ${\bar s}_b > 0$. 
In this case, in accordance with (\ref{subsN}), the substitution rate increases with ${\bar s}_b$ and approaches $E_0[k_b]$. 
For the larger population of size $N=10^{11}$, clonal interference operates when the mean beneficial effect exceeds ${\bar s}_b^*\approx 0.73 u_d$. Then, for ${\bar s}_b \ll {\bar s}_b^*$, only linked deleterious mutations matter resulting in substitution rates that coincide for both values of $N$. But for ${\bar s}_b^* \ll {\bar s}_b \ll u_d$, both deleterious and beneficial mutations affect the substitution rate and therefore it is smaller than the corresponding result for $N=10^8$. Finally for ${\bar s}_b \gg u_d$, only clonal interference is in effect and decreases the substitution rate from $E_0[k_b]$. 

Figure~\ref{fig5} shows the substitution rate for two population sizes as a function of mutation rate $u_d$ for a fixed fraction $p_b=u_b/u_d$ of beneficial mutations, and we find that it is a nonmonotonic function. For the population size $N=10^9$ for which clonal interference can be ignored, $E[k_b]$ initially increases linearly with $u_d$ due to the increased production of beneficial mutations and then decreases exponentially fast due to interference by deleterious mutations. For larger populations that satisfy $ 2 p_b N {\bar s}_b \ln (N {\bar s}_b) > e$, the condition $2 N p_b 
\ln (N {\bar s}_b) u_d^* e^{-u_d^*/{\bar s}_b}=1$ yields {\it two} solutions, 
\be
\frac{u_{d,1}^*}{{\bar s}_b}= -W_0 \left(\frac{-1}{2 p_b N {\bar s}_b \ln (N {\bar s}_b) } \right) 
\label{Lam0}
\ee
and
\be
\frac{u_{d,2}^*}{{\bar s}_b}= -W_{-1} \left(\frac{-1}{2 p_b N {\bar s}_b \ln (N {\bar s}_b) } \right) 
\label{Lam1}
\ee
where $W(x)$ is the Lambert W function \citep{Corless:1996}. This has the interesting consequence that a large population experiences clonal interference for a range of mutation rates given by $u_{d,1}^* < u_d < u_{d,2}^*$, but for $u_d < u_{d,1}^*$ and $u_d > u_{d,2}^*$, beneficial mutations fix sequentially. The data in Fig. 5 for $N=6 \times 10^{12}$ supports these conclusions as $E[k_b]$ is seen to be smaller than the corresponding result for $N=10^9$ for $0.02 {\bar s}_b \ll u_d \ll 4 {\bar s}_b$, and is independent of $N$ otherwise. 
  In the first case ($u_d < u_{d,1}^*$), clonal interference does not click due to the low production rate of beneficial mutations. But the origin of $u_{d,2}^*$ can be traced to the behavior of the fixation probability discussed in the preceding sections: when deleterious effects are extremely weak, only beneficial effects $ > u_d$ can fix and therefore with increasing  mutation rate, larger  beneficial effects are required for adaptation to proceed. But  the distribution of beneficial fitness effects is, in general, a decreasing function  \citep{Eyrewalker:2007} so that such large effects are rare. As a result, the large population again enters the sequential fixation regime where clonal interference is absent. This effect has been recently shown by \citet{Penisson:2017} when the beneficial and deleterious effects are nearly equal, and here we have provided a demonstration of the same in the limit $s_d \to 0$.


\noi{\bf Adaptation rate}

\noi We now address at what rate an asexual population adapts under the combined effect of interfering deleterious and beneficial mutations. The rate of adaptation is given by $R=E[k_b] \ln(1+E[s_b])$, where 
 \bea
 E[s_b] = \frac{\int_0^{S_b} ds~s P(s) e^{-I(s)} \rho(s)}{\int_0^{S_b} ds~P(s) e^{-I(s)} \rho(s)} 
 \label{seldef} 
 \eea
is the average selection coefficient of fixed beneficial mutation, and is smaller than one so that $R \approx E[k_b] E[s_b]$ . 

For $s_d \to 0$, proceeding in a manner similar to that for the substitution rate, we find the scaled rate of adaptation to be 
\be
\frac{R}{R_0}=\frac{1}{2}\int_{u_d/{\bar s_b}}^{\infty} dx~x^2 e^{-x} e^{-2 N u_b \ln(N {\bar s}_b x) (1+x^{-1}) e^{-x}} ~,
\label{adapdef}
\ee
that may be approximated as \citep{Wilke:2004}
\begin{subnumcases}{\frac{R}{R_0} \approx}
 \left(1+\frac{u_d}{{\bar s}_b}+\frac{u_d^2}{2 {\bar s}^2_b} \right) e^{-{u_d}/{{\bar s}_b}},&${\tilde A} e^{-{u_d}/{{\bar s}_b}} \ll 1$ \label{adaprates} \\
\frac{{(\ln {\tilde A})}^2}{2 {\tilde A}}-\left(\frac{u_d}{{\bar s}_b}\right)^2 \frac{e^{-{\tilde A} e^{-{u_d}/{{\bar s}_b}}}}{2 {\tilde A}},&${\tilde A} e^{-{u_d}/{{\bar s}_b}} \gg 1, {\bar s}_b \ll u_d$\label{adapratem} \\
\frac{{(\ln {\tilde A})}^2}{2 {\tilde A}},&${\tilde A} e^{-{u_d}/{{\bar s}_b}} \gg1, {\bar s}_b \gg u_d$\label{adapratel}
\end{subnumcases}
where, $R_0=4 N u_b {\bar s}_b^2$ is the adaptation rate in the absence of any interference.  As discussed for the substitution rate, the behavior of the adaptation rate may also be classified into four distinct regimes. 
Figure~\ref{fig6} shows the rate of adaptation as a function of population size for various mean beneficial effects, and we find that $R$ increases linearly with population size for $N < N^*$ and due to clonal interference, it grows slower than $N$  for $N > N^*$ where $N^*$ is the solution of ${\tilde A} e^{-u_d/{\bar s}_b}=1$. For ${\bar s}_b \gg u_d$, the adaptation rate is given by the corresponding result for lethal mutations where the effect of linked deleterious mutations can be neglected \citep{Wilke:2004}, but the interference by weakly deleterious mutations for ${\bar s}_b \ll u_d$ decreases the adaptation rate further. 

\noi{\bf Optimum mutation rate} 

\noi Like substitution rate, the rate of adaptation is also a nonmonotonic function of mutation rate, and is maximum at a mutation rate $u_{d}^{(opt)}$. The initial increase can be attributed to the increased production of beneficial mutations and the decay to the deleterious mutational load. For small populations $N \ll N^*$, from (\ref{adaprates}), we find that the (scaled) optimum mutation rate is a solution of the equation $2 + 2 x + x^2 - x^3 = 0$ which gives $u_{d}^{(opt)} \approx 2.27 {\bar s}_b$. The ratio $u_{d}^{(opt)}/{\bar s}_b$ is larger than one because the adaptation rate  increases linearly but the decrease is somewhat slower than an exponential (due to the factor in the bracket on the RHS of (\ref{adaprates})). For large populations $N \gg N^*$, the optimum mutation rate increases, albeit mildly, with $N$; these results are shown in the inset of Fig.~\ref{fig7}. \\

\noi{\bf \large{Fixation of Mutators in Adapting Asexual Populations}} 

\noi We now study the fixation of mutators when deleterious mutations are lethal and when they are extremely weak. In an adapting asexual population, mutator allele can spread because it is more likely to be  associated with beneficial mutations than the wild type \citep{Sniegowski:1997,Taddei:1997}. Since the mutator produces a beneficial mutant with probability $\approx U_b/U_d=u_b/u_d$ and this beneficial mutant fixes with probability $P$, the mutator can hitchhike to fixation if the probability $u_b P/u_d$ exceeds the neutral fixation probability $1/N$; this argument gives the critical population size $N_c$ above which the mutator can fix to be  $({u_b P}/{u_d})^{-1}$ \citep{Raynes:2018}. 


\noi{\bf Strongly deleterious mutations}

\noi When deleterious effects are strong, using (\ref{M2}) in the above argument for critical population size, we find that mutators with mutation rate $U_d > u_d+s_b$ can not fix in a finite population, and 
\be
N_c=\frac{u_d}{u_b} \frac{e^{u_d/s_d}}{2 (s_b+u_d-U_d)} ~,~U_d < u_d+s_b~,
\label{Ncstr}
\ee 
which shows that larger populations are required for the fixation of stronger mutators. 
Neglecting the reduction in fixation probability for large mutation rates, \citet{Raynes:2018} predicted the critical population size $N_c=(u_d/u_b) (2 s_b)^{-1}$ which is independent of mutator's mutation rate. For the parameters in panel B, Fig.~S2 of \citet{Raynes:2018}, the expression (\ref{Ncstr}) predicts the critical population size to be $555$ and $999$ for mutator of strength $100$ and $500$, respectively, in agreement with their simulation results. For $U_d=1000 u_d$, the condition $U_d < u_d+s_b$ is barely satisfied, but for mutator of strength $950$, we obtain $N_c \approx 9800$ from (\ref{Ncstr}), in close agreement with their simulations.

For exponentially distributed beneficial fitness effects, generalizing the clonal interference theory to mutators using (\ref{M2}), we find that the critical population size is determined by the following equation, 
\bea
\frac{u_d}{u_b} N_c^{-1} &=& \int_{U_d-u_d}^\infty ds_b \rho(s_b) P(s_b) e^{-N_c U_b \ln(N_c s_b) s_b^{-1} \int_{s_b}^\infty ds'  \rho(s') P(s')} \\
&=& 2 {\bar s}_b e^{-\mu} \int_{\upsilon}^\infty dx~e^{-x} (x-\upsilon) e^{-2 N_c U_b \ln(N_c {\bar s}_b x) e^{-\mu-x} (1+ \frac{1-\upsilon}{x})}~,
\label{Ncstrci}
\eea
where $\upsilon=(U_d-u_d)/{\bar s}_b$. Figure~\ref{fig7} shows the numerical solution of (\ref{Ncstrci}) for $N_c$, and we find that for strong mutators, clonal interference has a strong effect as it increases the critical population size substantially from when competition between beneficial mutations is neglected. 

\noi{\bf Weakly deleterious mutations} 

\noi For $s_d \to 0$, as the fixation probability of a mutant with mutation rate $U_d > s_b$ vanishes (see Fig.~\ref{fig2}), we immediately find that such a mutator can not fix in a finite population. But a population of size larger than $(u_d/u_b) (2 s_b)^{-1}$ is required for the fixation of a mutator with mutation rate $U_d < s_b$. Comparing this result with (\ref{Ncstr}) for lethal mutations, we find that somewhat larger populations are required for the spread of mutators when selection against deleterious mutations is strong. 

The effect of clonal interference can also be included for weakly deleterious mutations, and we have 
\be
\frac{u_d}{u_b} N_c^{-1}=2 {\bar s}_b \int_{U_d/{\bar s}_b}^\infty dx~x e^{-x} e^{-2 N_c U_b \ln (N_c {\bar s}_b x)  e^{-x} (1+ x^{-1})}~.
\label{Ncweaci}
\ee
A comparison of the results for $N_c$ found using (\ref{Ncstrci}) for strongly deleterious effects and the above equation suggests that the critical population size is mildly affected by the size of the deleterious effect, see Fig.~\ref{fig7}.  


\noi{\bf {Data availability}}

\noi The author states that all data necessary for confirming the conclusions presented in the article are represented fully within the article. \\

\noi{\bf \large{Discussion}}

\noi In this article, we focused on understanding how linked mutations affect the fixation of new beneficial mutations. Both deleterious \citep{Charlesworth:2012} and beneficial \citep{Sniegowski:2010} mutations are known to interfere with the evolutionary process in asexual populations. 

\noi {\bf Rate of adaptation in asexuals} 

\noi  When only deleterious mutations interfere, it is instructive to compare the extreme cases in which selection against deleterious mutations is very strong or very weak. In the former case, a beneficial mutation with  effect $s_b$ fixes with probability  (\ref{M2}) that decreases with increasing mutation rate $u_d$ from the classical result $2 s_b$, and approaches zero asymptotically. But in the latter case, while the corresponding fixation probability also decreases with $u_d$, the catastrophic reduction to zero occurs at a {\it finite} mutation rate that is equal to $s_b$ (see Fig.~\ref{fig2}). 
This transition in the fixation probability may be understood by the following back-of-the-envelope calculation: since the relative fitness of a beneficial mutant with $i$ deleterious mutations, each of deleterious effect $s_d$, is $(1+s_b) (1-s_d)^i e^{u_d}$, the mutant fixes with the probability $2 (s_b-i s_d+u_d)$ \citep{Haldane:1927}. But  the average number of deleterious mutations in the population is $\mu=u_d/s_d$ while the mutant can tolerate at most $\lambda=s_b/s_d$ deleterious mutations to avoid extinction. Thus, when $\mu < \lambda$, the fixation probability is $2 s_b$ (on replacing $i$ by its average $\mu$) and zero otherwise. For moderately sized deleterious effect, this transition has been observed, mainly numerically, in previous work \citep{Johnson:2002,Penisson:2017}, while here we have shown it analytically using the explicit expression (\ref{fullpexpr}) for the fixation probability.  

The above transition plays a key role in understanding the adaptation dynamics when both deleterious and beneficial mutations interfere with the fixation of a beneficial mutation.  
As summarized in Fig.~\ref{figphase}, when the scaled mutation rate $u_d/{\bar s}_b < 1$ (with ${\bar s}_b$ being the mean beneficial effect), 
interference by deleterious mutations can be ignored and we arrive at familiar results for the adaptation rate \citep{Orr:2000b,Wilke:2004,Sniegowski:2010}: in small populations where clonal interference is absent, it increases linearly with population size, whereas for larger populations, it increases logarithmically (see (\ref{adaprates}) and (\ref{adapratel})). 
However, when the scaled mutation rate exceeds one, linked deleterious mutations begin to interfere. Then for large mutation rates, the substitution rate becomes independent of the population size, unlike in the case where interference by deleterious mutations is neglected (cf. our Fig.~\ref{fig5} and Fig. 3 of \citet{Wilke:2004}). This means that in large populations, clonal interference disappears at high mutation rates \citep{Penisson:2017}. Because of the transition discussed above, at high mutation rates, only large beneficial effects stand a chance to spread, but such effects are rare and therefore competing beneficial mutations become unavailable. 


\noi{\bf Evolution of mutation rates in large asexual populations} 

\noi Although mutations drive the evolutionary process by creating novel genetic variation in a population, high mutation rates are generally not favored as mutators produce on an average more deleterious mutations. Thus in a well adapted population where beneficial mutations are not available, antimutators that lower the mutation rate are likely to be favored 
and the mutation rate will keep decreasing until the indirect selective advantage conferred due to reduced deleterious load becomes comparable to the strength of the genetic drift, $1/N$ \citep{Lynch:2010b,Lynch:2016}. 
But as the benefit of increasingly accurate replication diminishes, it follows that in an adapted population, mutation rate and population size are negatively correlated, and the quantitative relationship between them has been worked out for both strongly and weakly deleterious mutations 
\citep{James:2016,Jain:2017b}.  

On the other hand, in an adapting population, besides deleterious mutations, a mutator produces more beneficial mutations also, and in an asexual population, it 
may hitchhike to fixation with them thus increasing the mutation rate \citep{Sniegowski:1997,Taddei:1997}.  We first note that the mutation rate can not be arbitrarily high since the beneficial mutation carrying mutator may replace the resident wild type population if the former's fitness $(1+s_b) e^{-U_d}$ exceeds that of the latter, {\it viz.}, $e^{-u_d}$ (see (\ref{imax})). This general upper bound $U_d < s_b+u_d$ on the mutation rate gets further sharpened for very weakly deleterious mutations where mutator allele with mutation rate $U_d < s_b$ can only fix, refer (\ref{fullpexpr}). 
For mutation rates below these upper bounds, the (indirect) advantage offered by a mutator due to increased beneficial production rate, however, decreases with increasing mutator strength, 
and therefore a strong mutator may overcome the loss due to  random genetic drift in sufficiently large populations \citep{Raynes:2018}. 
Thus in an adapting population, the mutation rate and population size are positively correlated. 

From our analysis here based on this argument, we arrive at the following main conclusions: First, we find that the size of the deleterious effect does not significantly affect the critical population size needed for the fixation of mutators in adapting populations. 
In contrast, in the absence of beneficial mutations, the size of deleterious effect changes the quantitative relationship between population size and mutation rate (see (14) of \citet{James:2016}). Second, for both strongly and weakly deleterious effects, clonal interference increases the critical population size above which mutators are favored quite substantially (see Fig.~\ref{fig7}). This inflation in critical population size occurs because, as mentioned above, larger populations are needed for the fixation of stronger mutators. But beneficial mutations compete in large populations thus decreasing the fixation probability of a beneficial mutation-carrying mutator, and therefore even larger populations are required for the mutator to escape stochastic loss. 

A similar effect is also seen for the optimal mutation rate at which the rate of adaptation is maximum when the population enters the clonal interference regime. As the inset of Fig.~\ref{fig7} shows, the optimum mutation rate increases mildly with $N$; this is in contrast to the case when deleterious mutations are lethal where the optimum mutation rate remains independent of population size \citep{Orr:2000b}. We also note that when interference by deleterious mutations can be ignored, the population size above which clonal interference comes into play decreases with increasing mutation rate but it increases with mutation rate when deleterious mutations interfere, see Fig.~\ref{figphase}.

\noi {\bf Limitations and open questions} 

\noi Throughout this article, we worked in the parameter regime where the production rate of beneficial mutations is small ($N u_b < 1$) thereby neglecting the possibility of multiple-mutations in the same individual  \citep{Desai:2007a,Park:2007}. Extending the present work to include the combined effect of clonal interference and such multiple-mutations  \citep{Good:2012}, besides the interference by weak deleterious effects, would be interesting.

Here we focused our attention on the spread of beneficial mutations in a fully-linked genome, and an important direction for the future is to explore the effect of recombination on the results obtained in this work. As we have shown, a beneficial mutation can not fix in an asexual population beyond a critical mutation rate. 
But as the linked alleles dissociate over time in a  recombining population, 
it would be interesting to find how the fixation probability increases, particularly in the vicinity of the critical mutation rate,  with increasing recombination rate. More generally, for sufficiently high recombination rate, interference from linked  mutations are expected to disappear or get considerably weakened. But to what extent small recombination rates decrease the hindering effects of linked beneficial and deleterious mutations remains a question for the future. 


\noi {\bf Acknowledgements} 

\noi I thank Mark Kirkpatrick for helpful discussions during the early stages of this work. 

\clearpage

\appendix
\renewcommand{\thesection}{\Alph{section}}
\numberwithin{equation}{section}
%

\section{Quadratic approximation for fixation probability}
\label{app_quad}

\noi Let $p_i(t)$ denote the probability that a beneficial mutant arising at time $t$ in a wild type individual with $i$ deleterious mutations gets eventually fixed. If the mutant gives rise to $n$ offspring in the next generation with probability $\psi_i(n)$ and $j$ deleterious mutations occur in an offspring with  probability $M_{i\rightarrow i+j}$, the extinction probability $1-p_i(t)=\sum_{n=0}^{\infty} \psi_i (n) q_i(n,t)$ where $q_i(n,t)= \left[\sum_{j=0}^\infty M_{i \rightarrow i+j} ~ (1-p_{i+j}(t+1)) \right]^n $ is the probability that all the $n$ lineages go extinct \citep{Johnson:2002}. 
Here the mutation probability $M_{i \rightarrow i+j}=e^{-U_d} U_d^j/j!$, and the offspring number is also Poisson-distributed with mean $\omega_i$ equal to the fitness of the mutant relative to the average fitness of the resident population so that $\omega_i=(1+s_b) (1-s_d)^i/ e^{-u_d}$. For a slowly clicking ratchet, as the resident population remains at mutation-selection balance,  
the beneficial mutation arises and fixes in a stationary background and therefore the fixation probability is independent of time. 
Putting these pieces together results in the recursion equation (\ref{extprob1}) for the fixation probability. 
Furthermore, in order to have a nonzero fixation probability $p_i$, the fitness of the displacing population with at least $i$ deleterious mutations must exceed that of the resident population, {\it i.e.}, $(1+s_b) (1-s_d)^i e^{-U_d} > e^{-u_d}$ which means that $i \leq \lambda$, where $\lambda$ is given by (\ref{imaxf}). 

For $U_d \ll 1$, it is a good approximation to assume that an offspring suffers at most one deleterious mutation so that we may write $M_{i \rightarrow i+j} \approx (1-U_d) \delta_{j,0}+U_d \delta_{j,1}$. For the same reason, one may also ignore the contribution from those lineages in which more than one offspring undergoes mutation and write 
\bea
q_i(n) &\approx & \left[(1-U_d) (1-p_i) \right]^n+n U_d (1-p_{i+1}) \left[(1-U_d)(1-p_i)\right]^{n-1} \\
&\approx& 1- n p_i + \frac{n (n-1)}{2} p_i^2 +n U_d (p_i - p_{i+1})~,~
\eea
where we have used that the fixation probability $p_i$ is small because the relative fitness of the mutant $\omega_i \approx 1+s_b -i s_d +u_d$ is close to one when $s_b, s_d, u_d \ll 1$. Using these approximations in (\ref{extprob1}) and (\ref{imaxf}) leads to (\ref{receqn}) and (\ref{imax}), respectively, in the main text.


\section{Fixation probability for $U_d/s_b \to 0$}
\label{app_pert}

When $\lambda \gg 1$ but $\mu \ll \lambda$, the ratios $u_d/s_b, U_d/s_b \to 0$. We therefore expand  $p_i$ in a power series in the small parameter $u_d/s_b$ and write $p_i \approx p_i^{(0)}+(u_d/s_b) p_i^{(1)}$.  
Substituting this in (\ref{receqn2}), and keeping terms to linear order in $u_d/s_b$, we obtain
\bea
p_i &\approx& 2 (s_b -i s_d)+2 u_d - \frac{2 s_d U_d}{s_b-i s_d}~,~i < s_b/s_d \\
&\approx& 2 (s_b -i s_d)+2 u_d - \frac{2 s_d U_d}{s_b}~,~
\label{app_pert1}
\eea
where the last expression follows on using that for $U_d \ll s_b$, the mean of the frequency distribution $f_i$ is smaller than $\lambda$ and therefore fitness classes that contribute to the sum (\ref{fptot}) for total fixation probability are smaller than $\lambda$ ($i \ll \lambda$). If now $\mu \ll 1$, we may write the total fixation probability as $P \approx f_0 p_0+f_1 p_1$ with $f_1=\mu, f_0=1-f_1$ and arrive at (\ref{pert}). For $\mu \gg 1$, we can approximate the frequency distribution $f_i$ by a Gaussian distribution with mean and variance $\mu$. On carrying out the integral $P=\int_0^{\lambda} di f_i p_i$, we again obtain (\ref{pert}). 

\section{Fixation probability for finite $U_d/s_b$}
\label{app_finp}

To find the fixation probability of a beneficial mutant for $\lambda \gg 1$ and finite $\mu/\lambda$, we essentially follow the treatment in \citet{Jain:2017b}.  
We first note that when the background mutation $i \sim \lambda$, the second term on the left-hand side (LHS) of (\ref{receqn2}) may be ignored. Then the solution $p_{i,>}$ that obeys $p_{i,>}^2=2 U_d p_{i+1,>}$ is simply given by 
\be
p_{i,>}= 2 U_d \left(\frac{s_d}{U_d}\right)^{2^{i+1-\lambda}}~,~ 
\ee
where we have used that $p_{\lambda-1}=2 s_d$. 
We now write $p_i = p_{i, < }+ p_{i, >}$  in (\ref{receqn2}) to obtain
\bea
p_i \left[ p_{i,<} - 2 s_d (\lambda-i) \right] &=& 2 U_d p_{i+1,<}- p_{i,>} p_{i,<} \\
&\approx& 2 U_d (p_{i+1,<}- p_{i,<}) ~,~ \label{app_diff}
\eea
where the last expression follows on using that $p_{i,>} \stackrel{i \ll \lambda}{\longrightarrow} 2 U_d$. If we now set the LHS of (\ref{app_diff}) to zero, we get $p_{i,<} \approx 2 s_d (\lambda-i)$ 
which is a good approximation for small $s_d U_d$ or large $\mu$, as can be seen by using the approximate solution $p_{i,<}$ on the RHS of (\ref{app_diff}). 
Adding the above expressions for $p_{i,<}$ and $p_{i, >}$ yields (\ref{pexpr}). 

To find the total fixation probability, as before, we approximate $f_i$ by a Gaussian distribution with mean and variance equal to $\mu$. The integral over $p_{i,<}$ is exactly doable and gives the first two terms on the RHS of (\ref{fullpexpr}). To handle the term containing $p_{i,>}$, we approximate it by $2 U_d \Theta(i-i_\times)$ where $\Theta(x)$ is the Heaviside theta function and $p_{i_\times,>}=U_d$ which yields
\be
i_\times = \lambda - \frac{1}{\ln 2} \ln \left(\frac{\ln (U_d/s_d)}{\ln \sqrt{2}} \right)~,~
\ee
and the integral over the Gaussian distribution leads to the last term on the RHS of (\ref{fullpexpr}).

\clearpage

\begin{figure} 
\begin{center}  
\includegraphics[width=1\textwidth,angle=0,keepaspectratio=true]{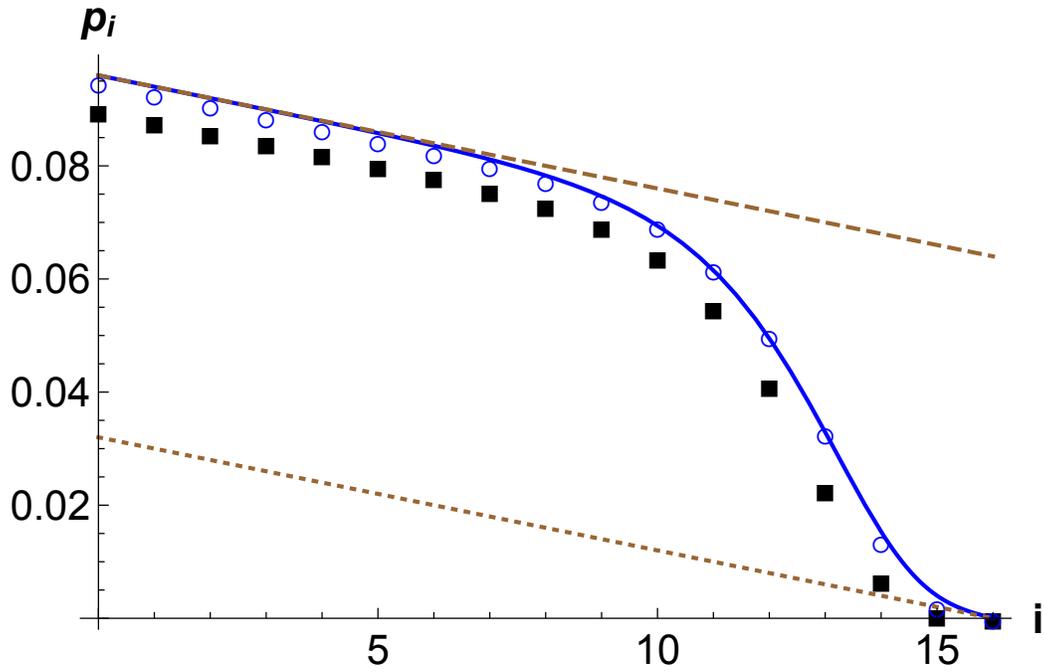}
\end{center}
\caption{Fixation probability $p_i$ as a function of background mutations  for large $\lambda$ and $\mu$. The points are obtained by numerically iterating the exact equation (\ref{extprob1}) (squares) and the quadratic-approximation (\ref{receqn}) (circles). The upper and lower bounds (dashed lines) given by (\ref{bdpi}) and the analytical expression (\ref{pexpr}) (solid line) are also shown. The parameters are $s_d=10^{-3}, u_d=8 s_d, U_d=4 u_d, s_b=0.04$ so that $\lambda=16, \mu=8$.}
\label{fig1}
\end{figure}

\clearpage
\begin{figure} 
\begin{center}  
\includegraphics[width=1\textwidth,angle=0,keepaspectratio=true]{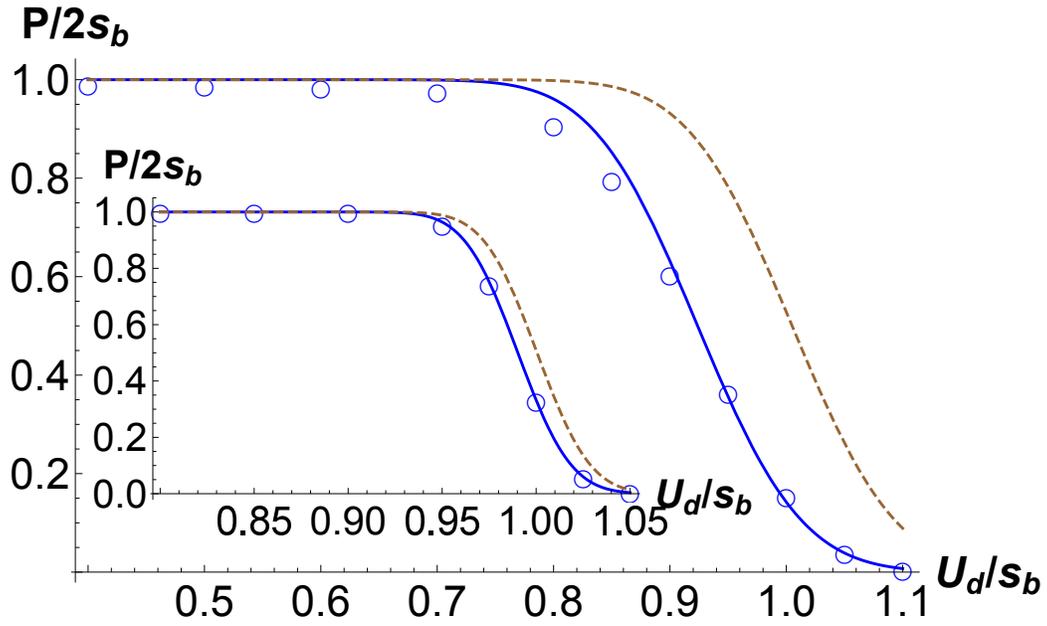}
\end{center}
\caption{Variation of total fixation probability $P$ with the mutation rate $U_d$ for large $\lambda$ and $\mu$. The data shown in circles is  obtained by numerically iterating  (\ref{receqn}) while the dashed and solid lines, respectively, show the  upper bound (\ref{pbd}) and the analytical expression (\ref{fullpexpr}). The parameters are $u_d=8 \times 10^{-3}, s_b=0.04$ and $s_d=10^{-3}$ (main), $10^{-4}$ (inset).}
\label{fig2}
\end{figure}

\clearpage
\begin{figure} 
\begin{center}  
\includegraphics[width=1\textwidth,angle=0,keepaspectratio=true]{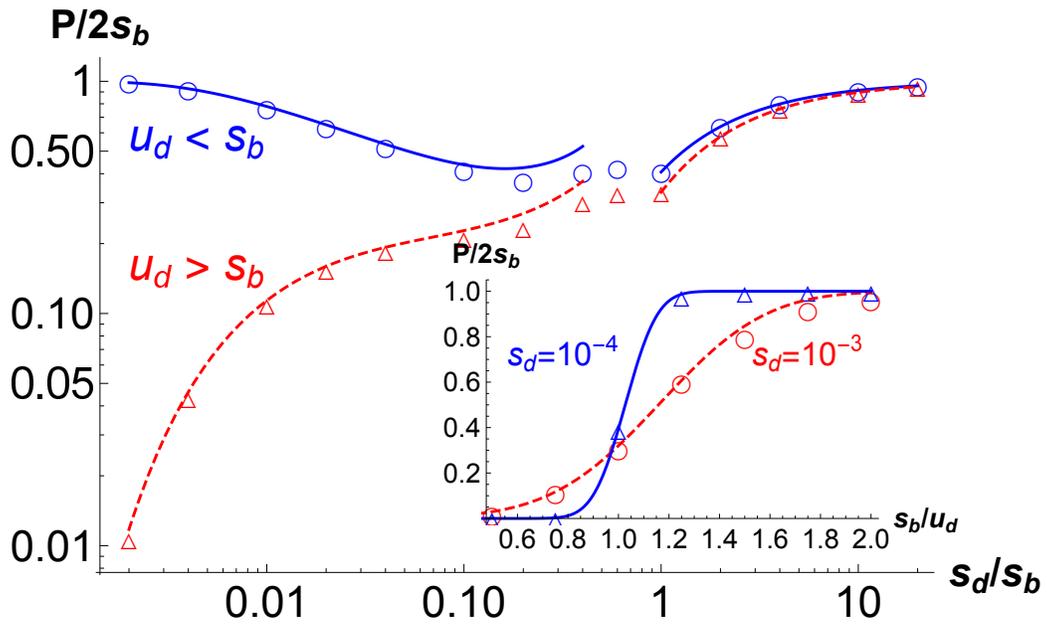}
\end{center}
\caption{Variation of total fixation probability $P$ with deleterious effect $s_d$ for  $U_d=u_d=0.9 s_b, 1.1 s_b$ and $s_b=0.005$ (main), and with beneficial effect $s_b$ for $U_d=u_d=0.01$ (inset). In both figures, the numerical solution of (\ref{receqn}) is shown by points, and the analytical result (\ref{fullpexpr}) by lines.}
\label{fig3}
\end{figure}

\clearpage
\begin{figure} 
\begin{center}  
\includegraphics[width=1\textwidth,angle=0,keepaspectratio=true]{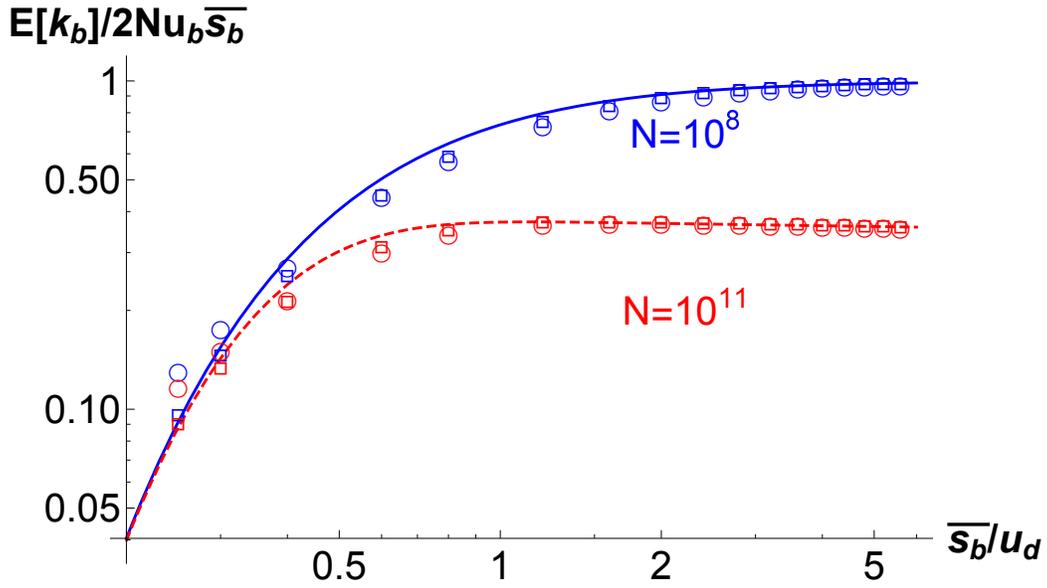}
\end{center}
\caption{Substitution rate $E[k_b]$ as a function of mean beneficial effect ${\bar s}_b$  for $u_d =5 \times 10^{-3}$ and  $u_b=10^{-12}$. The solid and dashed lines show the substitution rate (\ref{Eks00}) for $s_d \to 0$, while the points show the results obtained using (\ref{fullpexpr}) in (\ref{subsdef}) for $s_d=10^{-3}$ (circles) and $5 \times 10^{-4}$ (squares). For $N=10^{11}$, clonal interference occurs for ${\bar s}_b/u_d > 0.73$, while it remains absent for all ${\bar s}_b$ for $N=10^8$.}
\label{fig4}
\end{figure}

\clearpage
\begin{figure} 
\begin{center}  
\includegraphics[width=1\textwidth,angle=0,keepaspectratio=true]{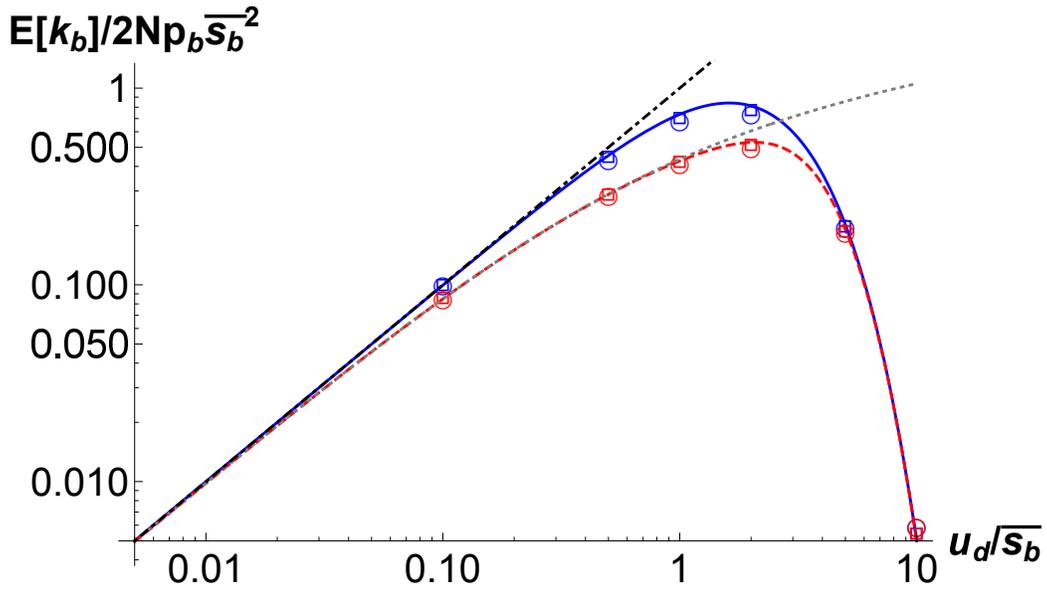}
\end{center}
\caption{Substitution rate as a function of mutation rate for $N=10^9$ (solid) and $6 \times 10^{12}$ (dashed). The dot-dashed curve and dotted curve, respectively, show the substitution rate when there is no interference and when only clonal interference occurs. The other parameters are  ${\bar s}_b= 10^{-2}, p_b= 10^{-12}$ for $s_d=10^{-3}$ (circles) and $5 \times 10^{-4}$ (squares). For $N=6 \times 10^{12}$, (\ref{Lam0}) and (\ref{Lam1}) predict that clonal interference occurs for $ 0.63 < u_d/{\bar s}_b < 1.49$.}
\label{fig5}
\end{figure}

\clearpage
\begin{figure} 
\begin{center}  
\includegraphics[width=1\textwidth,angle=0,keepaspectratio=true]{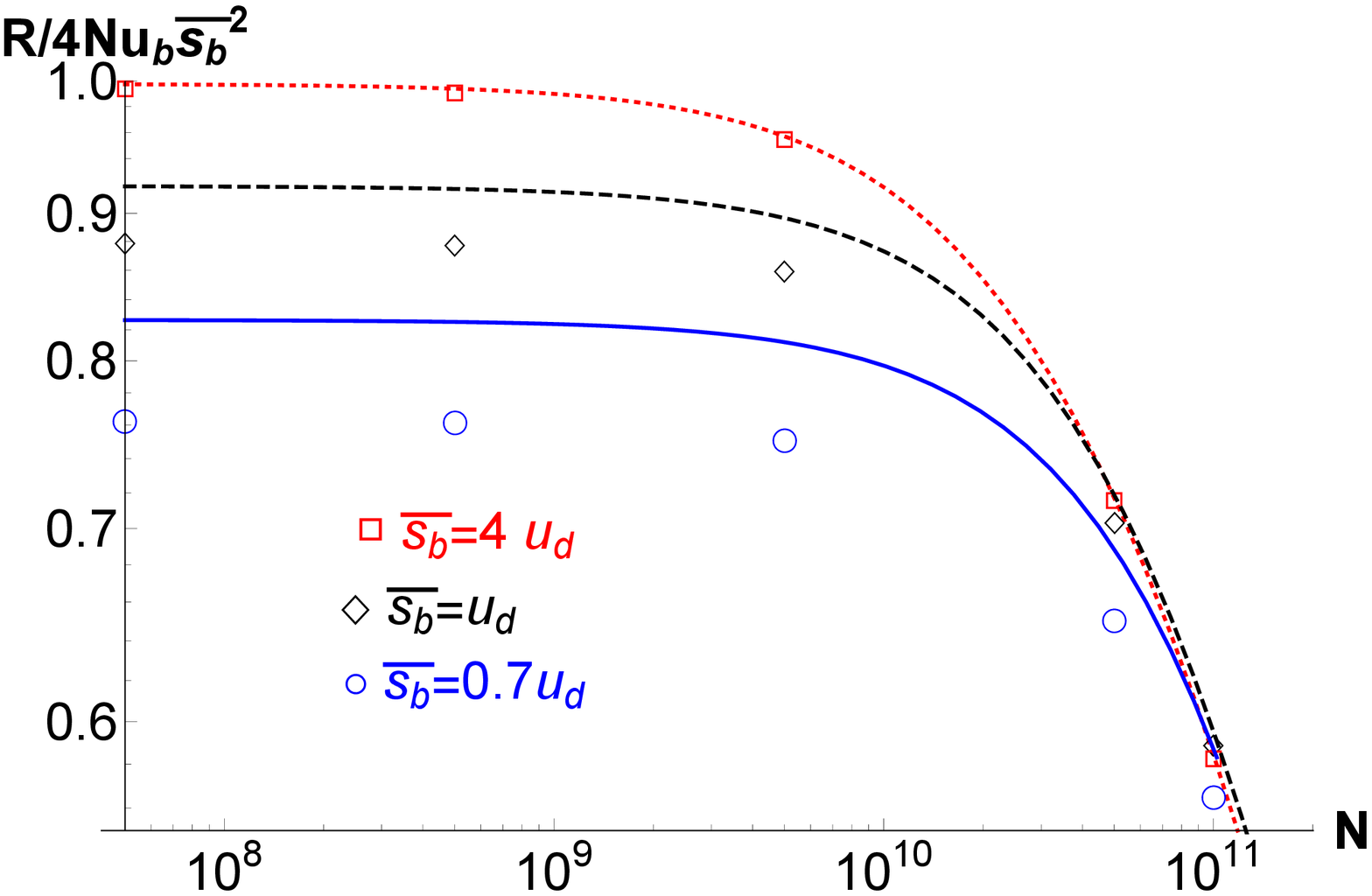}
\end{center}
\caption{Relative adaptation rate as a function of  population size $N$  for $u_d =5 \times 10^{-3}, u_b=10^{-12}$ and $s_d=5 \times 10^{-4}$ (points) for various ${\bar s}_b$. The lines show the result obtained using (\ref{adapdef}) for $s_d \to 0$. The population size above which clonal interference operates is found to be $3.2 \times 10^{10}, 6.9 \times 10^{10}, 1.1\times 10^{11}$ for  ${\bar s}_b/u_d=4, 1, 0.7$, respectively.}
\label{fig6}
\end{figure}

\clearpage
\begin{figure} 
\begin{center}  
\includegraphics[width=1\textwidth,angle=0,keepaspectratio=true]{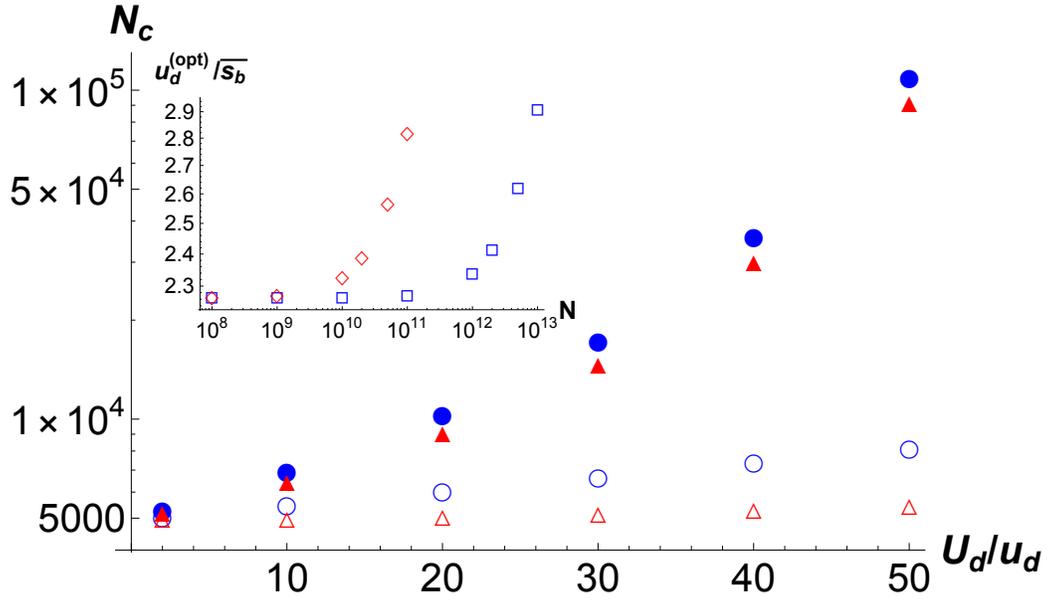}
\end{center}
\caption{Main: Critical population size $N_c$ above which mutators are favored when clonal interference is taken into account (filled symbols) and on neglecting it (open symbols). The data for strongly deleterious mutations (circles, $s_d=0.1$) and weakly deleterious mutations (triangles, $s_d \to 0$) are obtained by numerically solving (\ref{Ncstrci}) and (\ref{Ncweaci}), respectively. The parameters are $u_d=10^{-4}, u_b=10^{-6}, {\bar s}_b= 0.01$. Inset: Scaled mutation rate at which the rate of adaptation (\ref{adapdef}) for $s_d \to 0$ is maximum for  ${\bar s}_b= 0.01$ for the fraction of beneficial mutations $p_b=10^{-10}$ (diamonds) and $10^{-12}$ (square).}
\label{fig7}
\end{figure}

\clearpage
\begin{figure} 
\begin{center}  
\includegraphics[width=1\textwidth,angle=0,keepaspectratio=true]{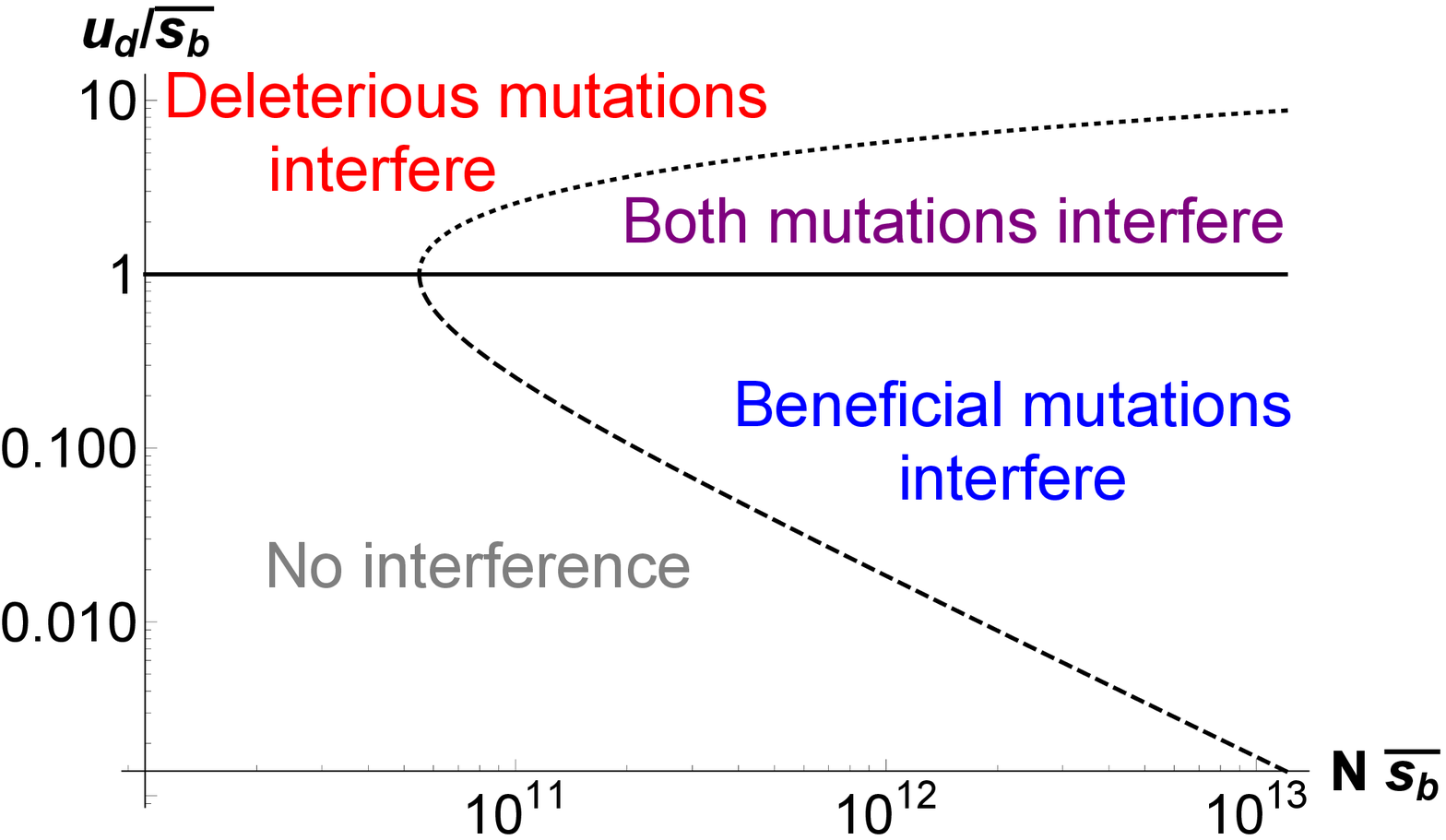}
\end{center}
\caption{Regions in the space of (scaled) population size and (scaled) mutation rate where linked mutations interfere when deleterious mutations are weak. The dashed and dotted lines, respectively, show (\ref{Lam0}) and (\ref{Lam1}) with beneficial mutation  rate $u_b=10^{-12}$, and the solid horizontal line separates the regimes where interference by deleterious mutations can be neglected and where it is effective.}
\label{figphase}
\end{figure}

\clearpage


\clearpage


\begin{center}
{\huge Supporting Information} 
\end{center}

\setcounter{figure}{0}
\setcounter{section}{0}

\setcounter{page}{1}
\makeatletter 
\renewcommand{\thefigure}{S\@arabic\c@figure} 
\renewcommand{\thesection}{S\@arabic\c@section} 
\renewcommand{\thetable}{S\@arabic\c@table}

\section{Stochastic simulations}

To test the theory developed in the main text, we simulated a population of $N$ haploid asexuals in which an  individual carrying $k_d$ deleterious and $k_b$ beneficial mutations was chosen to replicate with a probability proportional to its fitness  $(1-s_d)^{k_d} (1+s_b)^{k_b}~,~0 < s_b, s_d < 1$. The beneficial fitness effect $s_b$ was chosen from a truncated exponential distribution with mean ${\bar s}_b$ and maximum beneficial effect equal to $0.1$. Deleterious and beneficial mutations chosen from Poisson distribution with mean $u_d$ and $u_b$, respectively, were then introduced. Back mutations were ignored so that the number of beneficial and deleterious mutations increase with time. These steps were performed starting from a resident population at mutation-selection equilibrium with zero beneficial mutations, and each simulation was run until the time $T_{MR}$ when the Muller's ratchet clicked for the first time. 
For the background population to stay in equilibrium, we require the Muller's ratchet to click slowly and therefore the population size $N \gg s_d^{-1} e^{u_d/s_d}$ is required. 
Thus for weakly deleterious mutations, very large populations need to be simulated (e.g., for $s_d=2 \times 10^{-3}, u_d = 5 s_d$, a population of size $> 10^5$ is needed). With the available computational resources, we have been able to simulate population sizes up to $2 \times 10^5$ only.  

\begin{figure} 
\begin{center}  
\includegraphics[width=0.75\textwidth,angle=0,keepaspectratio=true]{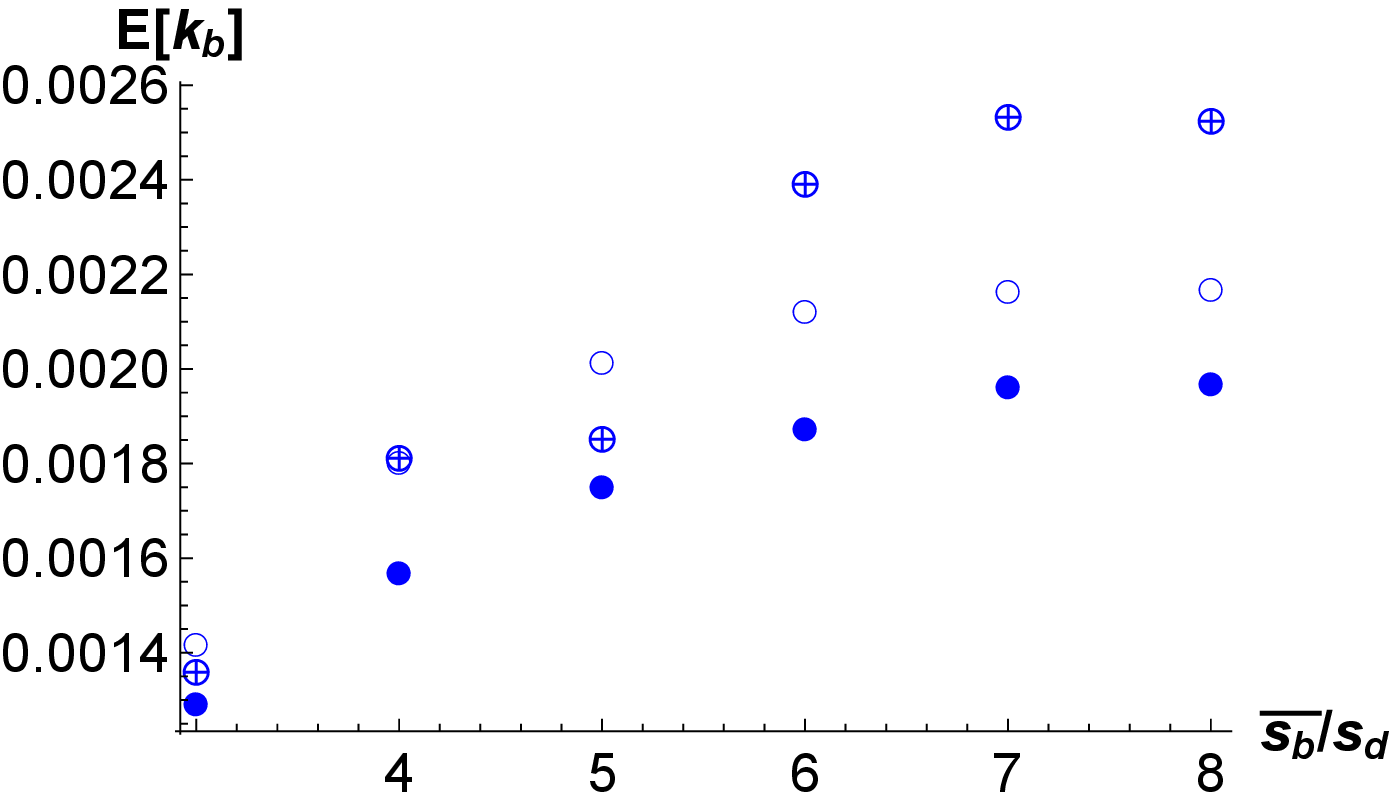}         
\includegraphics[width=0.75\textwidth,angle=0,keepaspectratio=true]{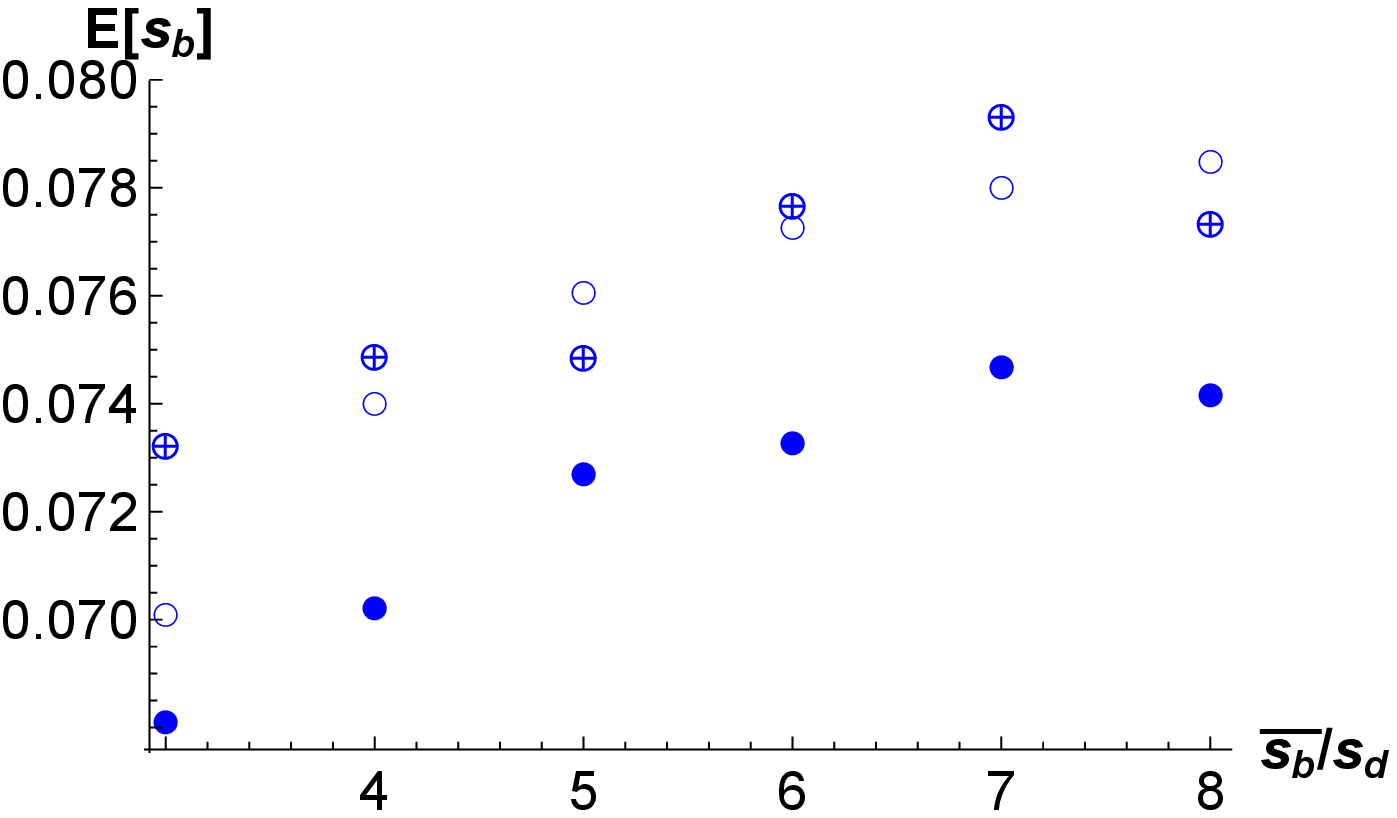}         
\end{center}
\caption{Expected substitution rate $E[k_b]$ and average fixed selection coefficient $E[s_b]$  for $N=2 \times 10^5$ when the average beneficial effect ${\bar s}_b$ is varied. The other parameters are $s_d=0.01, u_d=5 s_d, u_b=5 \times 10^{-7}$. The data from numerical simulations ($\bullet$) and numerical integration of (\ref{subsdef}) and (\ref{seldef}) using quadratic approximation (\ref{receqn}) ($\oplus$) and analytical expression (\ref{fullpexpr}) ($\circ$) is shown.}
\label{Sfig1}
\end{figure}

In each simulation run, the number of beneficial mutations ${\cal K}_b$ carried by maximum number of individuals in the population and the selection coefficient ${\cal S}_b$ averaged over the population were recorded. The average substitution rate $E[k_b]$ and the average selection coefficient $E[s_b]$ were then found by averaging ${\cal K}_b/T_{MR}$ and ${\cal S}_b$, respectively, over $10^3$ independent runs of the model described above.  
We also calculated these quantities using (\ref{subsdef}) and (\ref{seldef}) where the fixation probabilities were found numerically by solving the recursion equation (\ref{receqn}) and also using the analytical result (\ref{fullpexpr}). 
Figure~\ref{Sfig1} shows our results for the average substitution rate and average selection coefficient fixed using these three methods. We find that the results from stochastic simulations are consistently lower than those obtained using  (\ref{subsdef}) and (\ref{seldef}), possibly, because the simulation data is recorded until the Muller's ratchet clicks but the population is not in mutation-selection equilibrium close to $T_{MR}$. We also find that the results obtained using (\ref{receqn}) and (\ref{fullpexpr}) are also in quite good agreement  supporting the accuracy of our analytical predictions in the main text. 

\end{document}